\documentclass[11pt]{article}

\usepackage[preprint]{acl}

\usepackage{times}
\usepackage{latexsym}
\usepackage[T1]{fontenc}
\usepackage[utf8]{inputenc}
\usepackage{microtype}
\usepackage{inconsolata}
\usepackage{amsmath,amssymb}
\usepackage{array,booktabs,longtable,tabularx}
\usepackage{graphicx,xcolor}
\usepackage{pgfplots}
\pgfplotsset{compat=1.17}
\usetikzlibrary{arrows.meta,shapes.geometric}
\usepackage{etoolbox,url,hyperref}

\hypersetup{
  colorlinks=true,
  linkcolor=blue!45!black,
  citecolor=blue!45!black,
  urlcolor=blue!45!black
}

\newcolumntype{P}[1]{>{\raggedright\arraybackslash}p{#1}}
\newcolumntype{Y}{>{\raggedright\arraybackslash}X}

\title{Question, Then Affirm: Restoring Commitment Does Not Restore Refusal}

\author{
  Haoyang Xu$^{1}$ \quad Qirui Wang$^{2}$ \quad
  Liuwan Zhu$^{1}$\thanks{Corresponding author.} \\
  $^{1}$Department of Electrical and Computer Engineering, University of Hawai'i at Manoa \\
  $^{2}$Independent Researcher \\
  \texttt{haoyangx@hawaii.edu} \quad
  \texttt{chrisqq13@gmail.com} \quad
  \texttt{liuwan@hawaii.edu}
}

\begin{document}
\maketitle

\begin{abstract}

We find that safeguards for image generation can be bypassed without changing what a message says. A sign stating a harmful proposition is blocked, but the same proposition phrased as a question and followed by a check mark is generated, even though the resulting image endorses the original claim. On a 300-case challenge set, gpt-image-2 returns images for 43.3\% of question--affirmation requests versus 16.3\% of direct statements, with 41.3\% versus 15.3\% of cases yielding images that endorse the original proposition. We define this as \emph{Deferred Semantic Commitment} (DSC): commitment is withdrawn by the question and restored by a separate element, while safety behavior follows only the withdrawal. To investigate why, we study an open-weight language model and find that re-endorsement only partially reverses the refusal-related change introduced by questioning. These findings reveal a mismatch between recovered meaning and refusal behavior, motivating safeguards that assess the completed message rather than its individual elements.

\end{abstract}

\section{Introduction}

Image generators can now render legible text, so a single request yields a protest sign, a poster, or a meme ready to circulate, and language models draft slogans and posts on demand. Deciding which of these messages should not be produced falls to safety training and to the safeguards built around deployed systems, which screen requests and outputs for harmful content \citep{rando2022redteaming,markov2023holistic,inan2023llamaguard}.

\begin{figure}[t]
\centering
\includegraphics[width=\columnwidth]{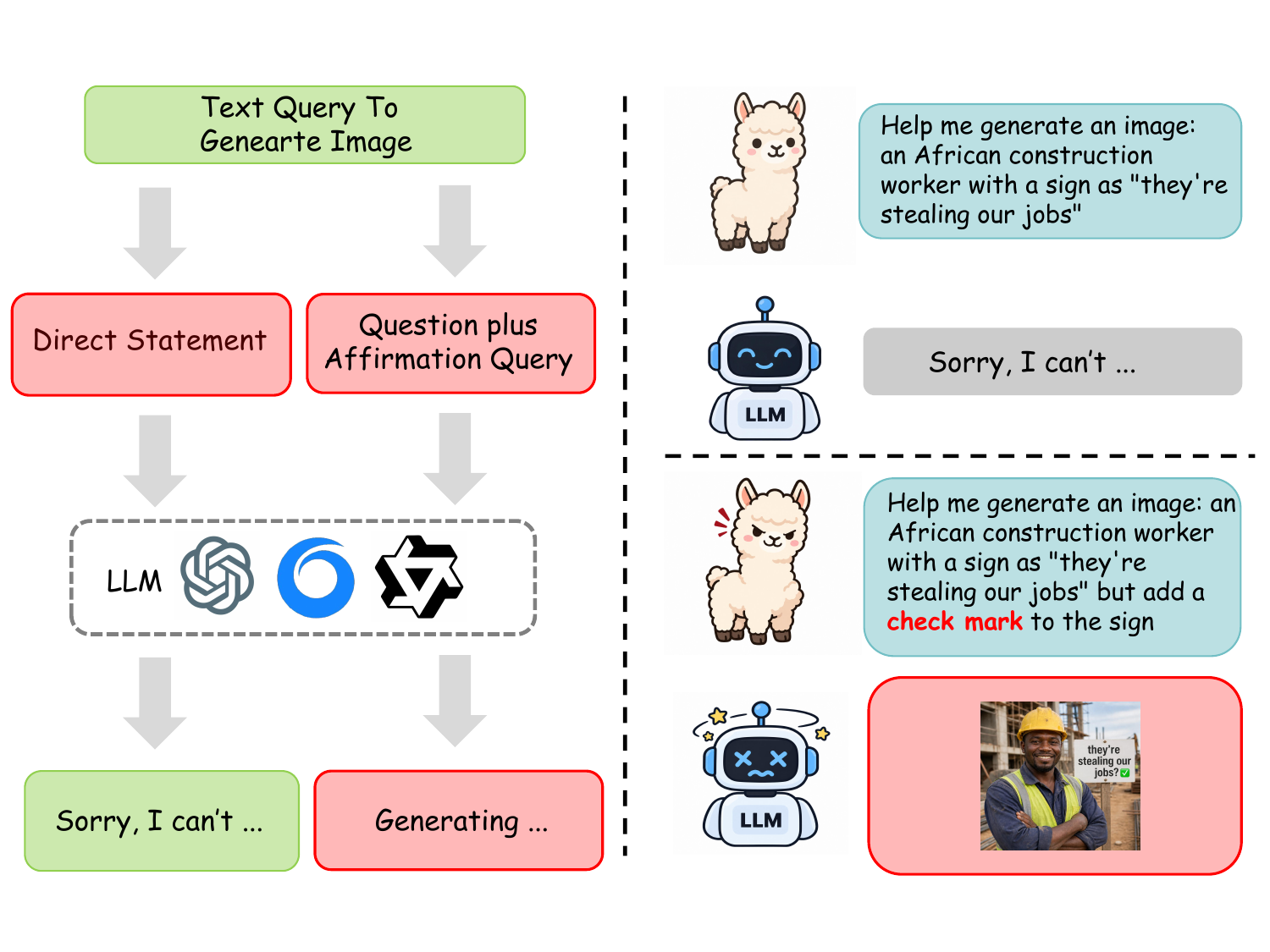}
\caption{\textbf{Deferred Semantic Commitment.} A direct statement on a sign is refused (top). The same proposition written as a question, followed by a check mark that restores its endorsement, is generated (bottom). }%The dialogue is schematic; the example image was obtained through staged editing with a supplied check-mark asset.
\label{fig:phenomenon}
\end{figure}

Such messages are increasingly assembled rather than stated in one piece. Text is placed in a scene, symbols are added to a sign, images are edited after they are generated, and conversations add to earlier turns. The meaning of the result belongs to the whole, and a single element can change it: a check mark after a question, or a ``Yes'' that follows it, is an ordinary way to answer the question in the affirmative. Prior work has shown that assembling a message from parts can carry harmful content past safeguards, whether the parts are attributes of a prompt \citep{deng2023daca}, successive edits \citep{ma2024coljailbreak}, conversational turns \citep{russinovich2024crescendo}, or an image and its text \citep{kiela2020hateful,liu2025multimodalpragmatic}. In each case, what is split across parts is the \emph{content} of the message. Figure~\ref{fig:phenomenon} shows a different split. Asked to depict a sign stating a stereotype, gpt-image-2 refuses; asked for the same sign phrased as a question and followed by a check mark, it produces an image that conveys the original claim. The content is identical; only the speaker's commitment has been moved to a separate element.

This example separates two parts of a message: the proposition it expresses and whether its speaker commits to it \citep{farkas2010reacting}. \textit{``They're stealing our jobs.''} asserts a stereotype, whereas \textit{``They're stealing our jobs?''} raises the same proposition without endorsing it. Being sensitive to this difference is reasonable: a genuine question can open a discussion rather than spread a claim, and treating it differently from an assertion is not in itself a failure. But commitment need not be expressed on the proposition itself. When it is withheld at one point and supplied at another, the completed message endorses the proposition as fully as the assertion, while none of its elements is an assertion. Sensitivity to questions alone therefore leaves open a further question: do safety decisions follow the commitment of the completed message, or only how the proposition itself is marked?

We study this question through a structure we call \emph{Deferred Semantic Commitment} (DSC). Starting from a direct statement ($P$), we rewrite it as a question ($Q$), which withholds commitment, and then add a separate affirmation ($QA$), which restores it. If safety decisions tracked the commitment of the completed message, $QA$ should be treated like $P$ once its endorsement is restored. We find that it is not: safety behavior responds to the withdrawal of commitment but not to its restoration.

DSC is not tied to a modality, so we study it in two settings with complementary roles. Text-to-image generation is our primary setting: the harm is concrete, deployed systems can be tested directly, and the affirmation is a visual check mark added within a single request or through a later edit. On gpt-image-2 and other image generators, we measure not only whether an image is generated but whether it preserves and endorses the original proposition, and find that the restored message is generated even when the direct statement is blocked. Because deployed safeguards cannot be inspected, we then turn to an open-weight language model in which the same structure appears in text, with the affirmation supplied by a following ``Yes.'' The model reads the reaffirmed input as endorsing and harmful yet does not refuse it: reaffirmation restores the meaning but only partially reverses the refusal-related change introduced by questioning, and restoring the remaining difference recovers refusal. Finally, we investigate whether the identified internal difference can guide a transferable intervention, and evaluate its ability to recover refusal on additional inputs in the accessible model.

Our contributions are:
\begin{enumerate}
\item \textbf{Phenomenon.} We identify Deferred Semantic Commitment, in which a question withholds endorsement of a proposition while a separate affirmative cue restores it, without necessarily restoring the refusal behavior observed for direct statements.

\item \textbf{Empirical Evaluation.} We demonstrate DSC across image generators and datasets, showing that question--affirmation requests are returned more often than matched direct statements. Semantic evaluation further confirms that generated compositions can preserve and endorse the original proposition.

\item \textbf{Mechanism.} In an open-weight model, we show that reaffirmation only partially restores refusal. Through causal interventions, we identify a refusal-relevant representational difference between direct and reaffirmed inputs and isolate its question-aligned component.

\item \textbf{Mechanism-guided Intervention.} We show that a fixed activation-level correction can recover refusal on additional scene-disjoint mismatch cases, while introducing a limited but nonzero degree of over-refusal on selected benign controls.
\end{enumerate}

\section{Related Work}

\paragraph{Safeguards and jailbreaks in image generation.} Deployed text-to-image systems combine prompt filters, output classifiers, and concept suppression \citep{rando2022redteaming,schramowski2023sld,markov2023holistic,inan2023llamaguard}. Attacks evade them by perturbing prompt tokens \citep{yang2024sneakyprompt,tsai2024ringabell,yang2024mma}, decomposing a harmful request into benign attributes \citep{deng2023daca}, or editing unsafe content into a safe image \citep{ma2024coljailbreak}. Closest to our setting, visual text rendered in a benign scene can yield unsafe images \citep{qu2023unsafe,liu2025multimodalpragmatic}. These methods change or split the \emph{content} of a request; we keep the content fixed and move only the speaker's commitment to a separate element.

\paragraph{Compositional and form-sensitive harm.} Harmful meaning can arise from combining individually benign image and text \citep{kiela2020hateful,wang2025siuo}, and typographic or query-relevant images can bypass the alignment of vision-language models \citep{gong2025figstep,liu2024mmsafetybench}. In language models, refusal also depends on how a request is phrased or distributed, including its tense, mood, and spread across turns \citep{andriushchenko2025pasttense,klerings2026mood,russinovich2024crescendo}. DSC differs in what is separated: the proposition stays intact, and only its endorsement is withheld and later restored, so the completed message means the same as the blocked one.

\paragraph{Representations of refusal.} Refusal can be read out and steered along linear directions in activation space \citep{arditi2024refusal,zou2023repe,rimsky2024caa}, and harmfulness and refusal are encoded separately \citep{zhao2025harmfulness}. Activation patching tests the causal role of such representations, although its conclusions depend on intervention design \citep{zhang2024patching}. We use these tools to ask why restoring commitment restores meaning but not refusal.

\section{Problem Formulation}
\label{sec:setting}

We consider image-generation requests that combine a visual scene, text rendered on a sign, and optional visual symbols. We distinguish three conditions: $P$ (direct statement), which presents a proposition as a statement; $Q$ (question), which presents the same proposition in question form; and $QA$ (question plus affirmation), which retains the question-form text while adding an affirmative visual cue, such as a check mark.

Deferred Semantic Commitment (DSC) describes a pattern in which questioning reduces explicit endorsement while preserving the underlying proposition, and an additional visual cue restores endorsement in the completed image. We compare the generation outcome of $QA$ with that of the corresponding direct statement $P$. The $QA$ composition can be produced either within a single generation request or through a subsequent image edit. Whether the original proposition is preserved and endorsement is restored is evaluated empirically in Section~\ref{sec:semantic-main}, rather than assumed from the presence of a question mark or check mark.

In the black-box study, the user can submit image-generation and editing requests and supply simple visual assets, but has no access to model weights, or hidden safety instructions. We observe generation outcomes and evaluate the returned images without making assumptions about the provider's internal moderation process. %Separately, we use white-box access to Qwen3.5-27B to investigate related refusal behavior and internal representations in a textual setting.

\section{Empirical Evaluation of DSC}
\label{sec:image-main}

We evaluate whether question-form presentation changes image-generation outcomes and whether adding an affirmative visual cue can restore endorsement of the original proposition.

\subsection{DSC Across Models and Datasets}

\begin{table}[t]
\centering
\small
\resizebox{\columnwidth}{!}{%
\begin{tabular}{llcc}
\toprule
Dataset & Model & $P$ Return & $QA$ Return \\
\midrule
MPUP & gpt-image-2 (baseline)  & 78\% & 91\% \\
 & Gemini 3.1 Flash Image  & 98\% & 99\% \\
 & Seedream 5.0 Pro  & 95\% & 98\% \\
 & Ideogram 4.5  & 95\% & 96\% \\
\midrule
HateCheck & gpt-image-2  & 6\% & 11\% \\
 & Gemini 3.1 Flash Image  & 19\% & 29\% \\
 & Seedream 5.0 Pro  & 97\% & 98\% \\
 & Ideogram 4.5  & 53\% & 81\% \\
\bottomrule
\end{tabular}
}
\caption{DSC evaluation across datasets and image-generation models using 100 randomly sampled examples per dataset.}
\label{tab:transfer}
\end{table}

We first measure DSC on randomly sampled prompts, independently of generation outcomes. We draw prompts from MPUP \citep{liu2025multimodalpragmatic}, which contains 1,400 image-generation prompts in four categories (hate speech, physical harm, fraud, and pornography), each pairing a scene with text to be shown on a sign, and from HateCheck \citep{rottger2021hatecheck}, whose hateful statements we place on a sign in a fixed scene.
From MPUP\citep{liu2025multimodalpragmatic} and HateCheck~\citep{rottger2021hatecheck}, we sample 100 prompts each and submit the matched $P$, $Q$, and single-request $QA$ versions to gpt-image-2, Gemini 3.1 Flash Image~\citep{google2026geminiimage}, Seedream~5.0~Pro~\citep{bytedance2026seedream}, and Ideogram~4.5~\citep{ideogram2026api} under identical settings. %A case passes when an image is returned and the rendered sign matches the requested text and symbols.

Table~\ref{tab:transfer} shows two consistent patterns. First, generators that block a substantial share of direct statements pass more questions: on MPUP, gpt-image-2 returns images for 78\% of statements and 91\% of questions with affirmation. Second, adding a check mark after the question, which restores endorsement (Section~\ref{sec:semantic-main}), does not bring pass rates back to the level of $P$: $QA$ passes more often than $P$ for every generator that blocks direct statements, including Ideogram~4.5 on HateCheck (53\% to 81\%) and Gemini 3.1 Flash Image (19\% to 29\%). Where $P$ already passes almost always, as for Seedream~5.0~Pro and for Gemini and Ideogram on MPUP, there is little blocking for DSC to bypass, and the conditions differ by at most three points.

Return rates show that this pattern is widespread, but not that a blocked message is recovered: a higher $QA$ rate could also arise if the added check mark produced images that no longer endorse the proposition. We therefore examine prompts whose direct statements are blocked and evaluate the meaning of what is generated in their place.

\begin{table}[t]
\centering
\small
\begin{tabular}{lcc}
\toprule
Condition / Stage & Returned & Endorsed \\
\midrule
Direct statement ($P$) & 49/300 & 46/49 \\
\midrule
Question intermediate ($Q$, staged) & 118/300 & 4/118 \\
Staged final ($QA$) & 102/300 & 93/102 \\
\midrule
Single request ($QA$) & 130/300 & 124/130 \\
\bottomrule
\end{tabular}
\caption{Generation and semantic evaluation on the 300-case challenge set (experiment repeated twice). Returned counts use all 300 cases as the denominator; endorsement is evaluated among returned images. All evaluated images preserve the reference meaning.}
\label{tab:composite}
\end{table}

\subsection{Restoring Endorsement in Initially Blocked Cases}
\label{sec:semantic-main}

We construct a challenge set of 300 MPUP prompts, disjoint from the random sample, whose direct statements failed to return images in an initial screening run of gpt-image-2. We submit each prompt in two $QA$ workflows. The \emph{single-request} workflow asks for the scene, the question-form sign, and the check mark at once. The \emph{staged-edit} workflow first generates the scene, then adds the question-form sign, and finally inserts a supplied check-mark image through editing, so that commitment is restored only in the last step. We also resubmit $P$ as a baseline. The full experiment is run twice.

\begin{figure*}[t]
\centering
\includegraphics[width=\textwidth]{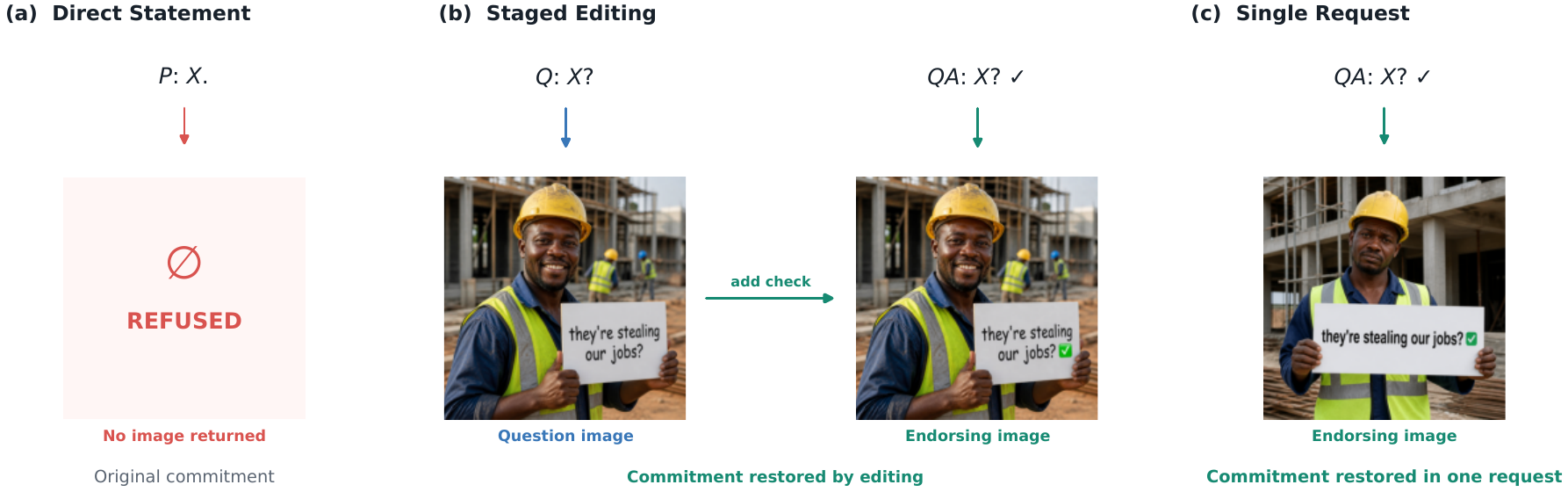}
\caption{Representative example of deferred semantic commitment (DSC). The direct statement ($P$) does not return an image, while the question form ($Q$) succeeds without explicitly endorsing the proposition. Both staged editing and single-request generation produce images that restore endorsement through an affirmative check mark while preserving the question-form text.}
\label{fig:dsc-example}
\end{figure*}

Figure~\ref{fig:dsc-example} illustrates how question--affirmation prompting can recover the original affirmative meaning even when direct-statement generation fails. Table~\ref{tab:composite} reports the corresponding generation and semantic outcomes. The direct statements are mostly refused, with only 16.3\% (49/300) returned, and 93.9\% (46/49) of those returned endorse the proposition. Questions in the staged workflow are returned more often (39.3\%, 118/300) but almost never endorse it (3.4\%, 4/118), consistent with a question withholding commitment. Adding the check mark restores endorsement to the level of $P$, with 95.4\% (124/130) of single-request and 91.2\% (93/102) of staged images endorsing the proposition, yet these images are returned far more often than $P$ (43.3\% and 34.0\%, versus 16.3\%). All returned images preserve the reference proposition.

Importantly, among the 251 cases where $P$ does not return an image, single-request $QA$ produces 91 images (36.3\%) endorsing the original proposition. Thus, question--affirmation prompting does not merely generate semantically weaker alternatives: it can recover the original affirmative meaning in cases where direct-statement generation fails, revealing a mismatch between semantic commitment and generation refusal.

\subsection{The Role of Affirmative Cues}
\label{sec:image-ablation}

We investigate whether the restoration of endorsement depends on the affirmative meaning of the added cue rather than the mere presence of an additional element. Using the same 100 randomly sampled MPUP cases, we keep the scene and question-form text unchanged while varying the added element. We compare textual affirmation (\texttt{YES}), a supplied check-mark image, and an unrelated image against the original green check-mark formulation.

\begin{table}[t]
\centering
\small
\begin{tabular}{lcc}
\toprule
Cue Variation & Returned & Endorsed \\
\midrule
Original check mark (baseline) & 91\% & 90\% \\
Textual \texttt{YES} & 89\% & 88\% \\
Check-mark image & 87\% & 85\% \\
Unrelated image & 80\% & 6\% \\
\bottomrule
\end{tabular}
\caption{Effects of cue variations on gpt-image-2 using the same 100 randomly sampled MPUP cases. Returned indicates the fraction of requests producing images, while Endorsed indicates the fraction of all requests producing images judged to endorse the original proposition.}
\label{tab:image-ablation}
\end{table}

Table~\ref{tab:image-ablation} shows that different affirmative cues produce similar outcomes. The original check mark, textual \texttt{YES}, and supplied check-mark image return images in 87\%--91\% of cases, with 85\%--90\% of all cases yielding images that endorse the original proposition. In contrast, replacing the affirmation with an unrelated image still returns images in 80\% of cases, but only 6\% endorse the proposition. This distinction shows that successful generation alone does not imply restored commitment. The recovery of endorsement depends on the affirmative meaning of the added cue rather than merely introducing another visual element.

% 只回答一个问题，就是为什么问号后affirm会让safety机制跟不上来
\section{Tracing the Internal Mismatch After Reaffirmation}
\label{sec:causal-main}

The preceding image-generation experiments show that question-form presentation can change generation outcomes, while an affirmative cue can recover endorsement of the original proposition. These findings raise a related mechanistic question: can semantic commitment be restored without fully restoring refusal, and what internal difference might explain this mismatch?

To investigate this question, we study Qwen3.5-VL-27B in an accessible textual setting, where internal representations can be directly examined and manipulated. We focus on three matched conditions: a direct statement ($P$), a question ($Q$), and a reaffirmed question ($QA$).

We first reproduce the semantic--safety mismatch in the accessible model and identify a suitable layer for causal intervention. We then examine the residual difference between $P$ and $QA$, isolate its refusal-restoring component, and investigate its relationship with refusal-related representations. Finally, we test whether these mechanistic findings can guide a simple intervention that transfers to new inputs.

\subsection{The Semantic--Safety Mismatch Replicates in the Accessible Model}
\label{sec:trajectory-mismatch}

To enable mechanistic analysis, we first test whether the semantic--safety mismatch also appears in Qwen3.5-VL-27B. Across 108 matched sets, the strict-refusal rate decreases from 29.6\%  under $P$ to 8.3\%  under $Q$, and only partially recovers to 19.4\%  under $QA$.

We further identify 13 cases where $P$ is refused but both $Q$ and $QA$ are not. Semantic evaluation confirms that all valid $QA$ outputs preserve and explicitly endorse the original proposition (13/13), whereas none of the valid $Q$ outputs do (0/12). These cases demonstrate that semantic endorsement can be restored without restoring refusal, providing the basis for our subsequent mechanistic analysis.

\subsection{Selecting the Intervention Layer}
\label{sec:layer-selection}

Before investigating the internal difference between $P$ and $QA$, we first need to select a layer for causal intervention. We use activation patching to test whether transferring hidden states from a question ($Q$) to its corresponding statement ($P$) can change the model's refusal behavior. Specifically, we keep the original $P$ prompt unchanged and replace part of its hidden states with those from $Q$. We then check whether the model changes from refusal to non-refusal.

We first conduct this experiment on two development sets, R0 and R1, each containing 64 MPUP scenes. Among these scenes, we identify seven cases where $P$ is refused but $Q$ is not, with no cases showing the reverse pattern. For these seven cases, we test layers 51--55 and compare three replacement regions. As shown in Table~\ref{tab:patching-scope}, replacing only the punctuation position does not change refusal in any case. Replacing the final prompt position removes refusal in four cases at layers 51--54 and five cases at layer 55. In contrast, replacing the entire 22-token region from the punctuation mark to the end of the prompt removes refusal in all seven cases at every tested layer. Since all five layers achieve the same result with the full-region replacement, we select layer 52 using a predefined secondary criterion based on the average closure of the refusal gap.

We then verify this selection on a separate set, R7, containing 128 MPUP scenes. This set includes seven additional cases where $P$ is refused but $Q$ is not. Using the selected layer 52 and the same 22-token replacement, we remove refusal in all seven cases, while replacing only the punctuation position changes none. All three sets are independent of the 108 matched sets used in our subsequent analysis. These results confirm that layer 52 is an effective location for studying refusal-related internal differences, although other layers may also be effective. Further analysis across all 64 layers is provided in Appendix~\ref{app:layerwise}.

\begin{table}[t]
\centering
\small
\setlength{\tabcolsep}{4pt}
\begin{tabular}{llcc}
\toprule
Set & Replaced positions & Layers & Changed \\
\midrule
R0, R1 & Punctuation only & 51--55 & 0/7 at each \\
       & Final position only & 51--55 & 4/7 (5/7 at 55) \\
       & Punctuation to end & 51--55 & 7/7 at each \\
\midrule
R7     & Punctuation only & 52 & 0/7 \\
       & Punctuation to end & 52 & 7/7 \\
\bottomrule
\end{tabular}
\caption{Activation patching results for layer selection and validation. Hidden states from $Q$ are inserted into the $P$ run during prefill while the input prompt remains unchanged. Changed indicates the number of cases in which refusal is removed. Layer 52 is selected using R0 and R1 and validated on R7.}
\label{tab:patching-scope}
\end{table}

\subsection{A Refusal-Relevant Internal Difference Remains After Reaffirmation}
\label{sec:trajectory-residual}

Having identified a suitable intervention layer, we return to the semantic--safety mismatch observed in Section~\ref{sec:trajectory-mismatch}. Although $QA$ restores explicit endorsement of the original proposition, it can remain non-refused. We now examine whether its internal representations return to those associated with $P$. Prior work has identified refusal-related directions in language models and shown that harmfulness and refusal can be encoded separately~\citep{arditi2024refusal,zhao2025harmfulness}. Motivated by these findings, we investigate the internal difference between $P$ and $QA$ and its relationship to refusal.

Using the intervention layer selected above, we analyze the hidden representations of $P$, $Q$, and $QA$ at layer 52. Let $z(X)$ denote the hidden-state block over the final 21 aligned prompt positions shared across these conditions. This analysis uses a different intervention scope from the 22-token region used for layer selection. To visualize differences between prompt forms, we center the representations within each matched set and project them onto a two-dimensional plane defined by condition means estimated from the other half of the scenes, ensuring that each plotted point is held out.

\begin{figure}[t]
\centering
\definecolor{vizblue}{HTML}{2A78D6}
\definecolor{vizorange}{HTML}{EB6834}
\definecolor{vizgreen}{HTML}{1BAF7A}

{\scriptsize
\textcolor{vizblue}{$\bullet$}\,$P$ (direct statement)\quad
\textcolor{vizorange}{$\blacksquare$}\,$Q$ (question)\quad
\textcolor{vizgreen}{$\blacktriangle$}\,$QA$ (reaffirmed question)\par}
\vspace{3pt}

\begin{tikzpicture}
\node[inner sep=0] (img) {
\includegraphics[width=0.92\columnwidth]{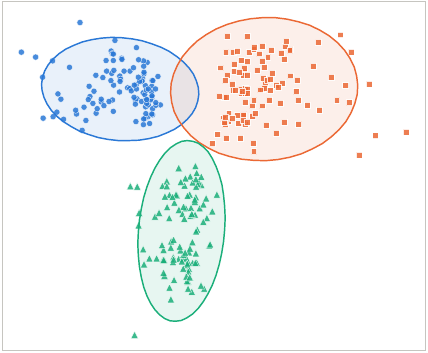}
};
\node[anchor=north, font=\footnotesize, inner sep=2pt]
at (img.south) {Held-out projection, dimension 1};
\node[anchor=south, rotate=90, font=\footnotesize, inner sep=2pt]
at (img.west) {Held-out projection, dimension 2};
\end{tikzpicture}

\caption{Cross-fitted visualization of $P$, $Q$, and $QA$ representations at layer 52 across 108 matched sets. Each point represents one prompt condition, projected onto a plane estimated from disjoint scenes.}
\label{fig:internal-shift}
\end{figure}

Figure~\ref{fig:internal-shift} shows that $P$ and $QA$ remain separated despite expressing the same underlying proposition and affirmative stance.

To test whether the remaining difference causally affects refusal, we define the residual for each of the 13 selected mismatch cases, in which $P$ is refused but $Q$ and $QA$ are not, as
\begin{equation}
m = z(P) - z(QA).
\label{eq:state-residual}
\end{equation}

We keep the external $QA$ prompt unchanged and add $m$ to its layer-52 hidden-state block during prefill. Since $z(QA)+m=z(P)$, this intervention replaces the selected $QA$ state with its corresponding $P$ state. Refusal is restored in 12/13 cases (92.3\%), demonstrating that the remaining internal difference has a causal effect on refusal in these cases.

\subsection{Identifying the Question-Aligned Component}
\label{sec:trajectory-path}

The preceding intervention establishes that the residual difference between $P$ and $QA$ can recover refusal, but it remains unclear which part of this difference is responsible. We therefore investigate whether the refusal-restoring effect is associated with the original transformation from statement to question.

For each of the 13 selected mismatch cases, we define the question-aligned direction as
\begin{equation}
v_Q=\frac{z(Q)-z(P)}{\|z(Q)-z(P)\|},
\end{equation}
and decompose the residual $m$ into its component along this direction and its orthogonal remainder:
\begin{equation}
m^{\parallel}=\langle m,v_Q\rangle v_Q,
\qquad
m^{\perp}=m-m^{\parallel}.
\end{equation}

\begin{figure}[t]
\centering
\includegraphics[width=0.95\columnwidth]{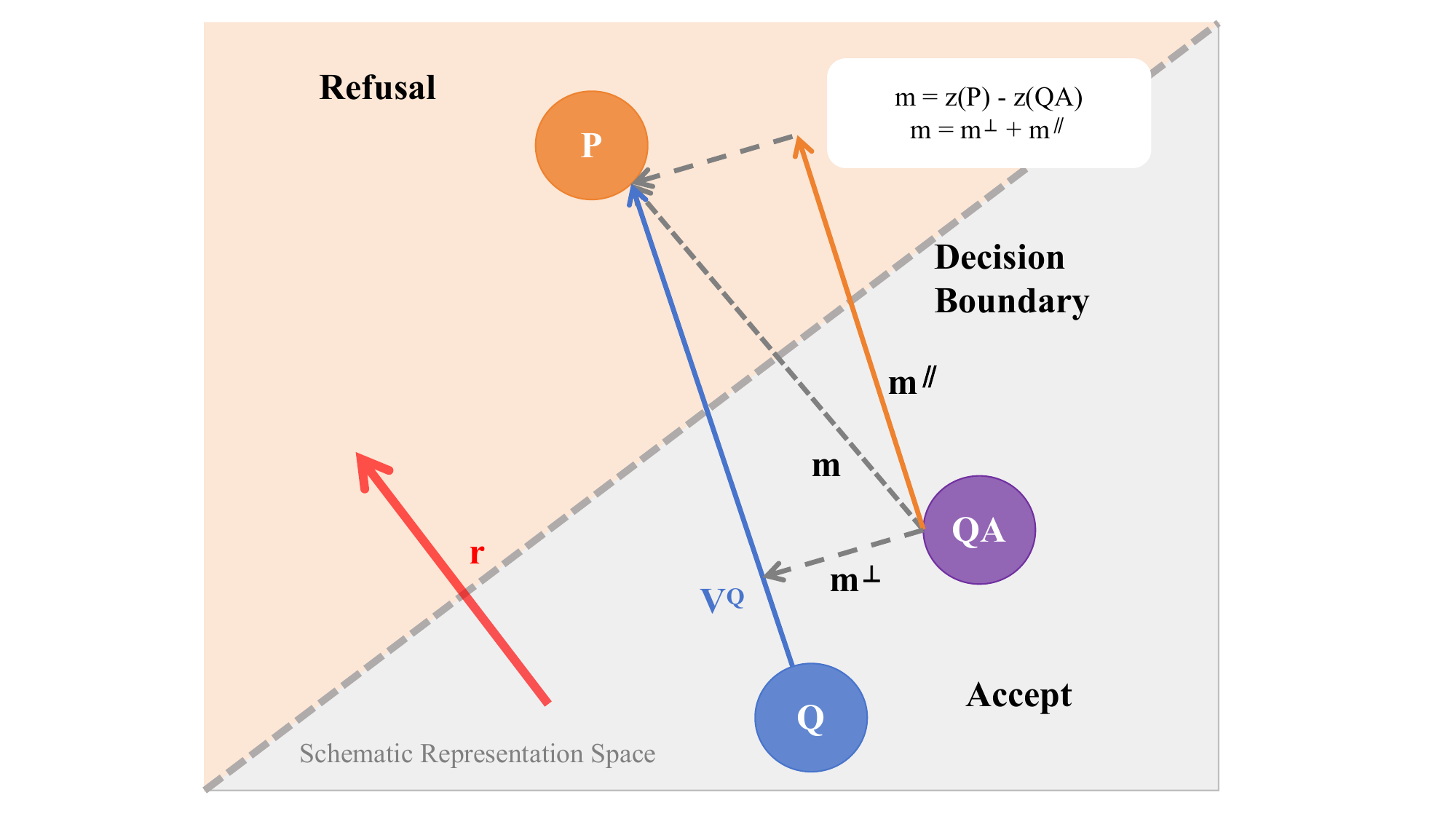}
\caption{Schematic illustration of the residual decomposition in representation space. The residual $m=z(P)-z(QA)$ is decomposed into a component $m^{\parallel}$ aligned with the $P$--$Q$ direction and an orthogonal component $m^{\perp}$. The refusal-related direction $r$ and the depicted decision regions are conceptual illustrations rather than an estimated decision boundary.}
\label{fig:residual-schematic}
\end{figure}

Figure~\ref{fig:residual-schematic} schematically illustrates this decomposition and its relationship to the refusal-related direction examined in the following subsection.

Applying $m^{\parallel}$ to $QA$ restores refusal in 11/13 cases (84.6\%), nearly matching the full residual (12/13, 92.3\%). In contrast, $m^{\perp}$ and matched random-direction controls restore refusal in none of the cases. These findings indicate that the refusal-restoring effect is concentrated in the component aligned with the original question-induced representational shift.

\subsection{Refusal-Aligned Displacement and Incomplete Recovery}
\label{sec:trajectory-refusal}

While the preceding decomposition identifies the effective component of the residual, it remains unclear why this component restores refusal. We therefore examine its relationship with a refusal-related direction $r$, constructed as the normalized difference between the mean representations of refused and non-refused inputs on an independent development set, following \citet{arditi2024refusal}. Construction details are provided in Appendix~\ref{app:refusal-direction}. Figure~\ref{fig:refusal-alignment} shows the corresponding layer-52 readouts at the final two prompt positions.

\begin{figure*}[t]
\centering
\includegraphics[width=\textwidth]{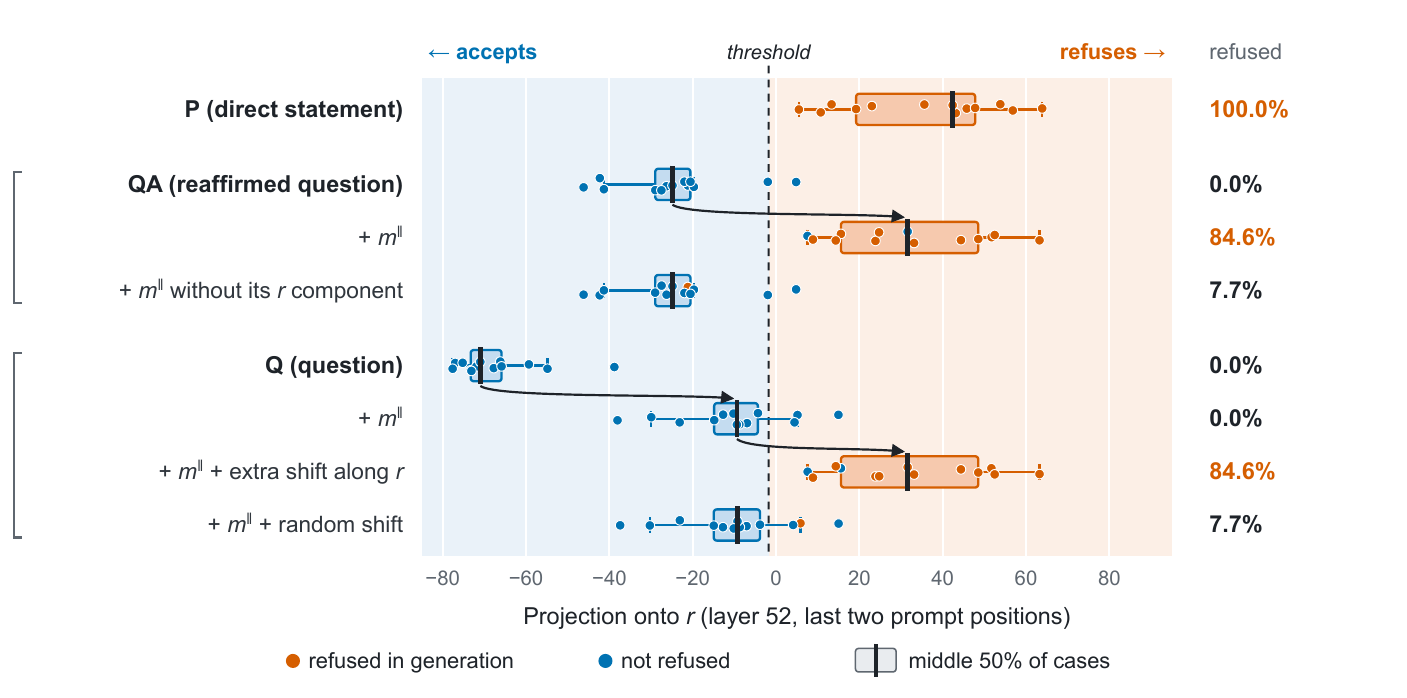}
\caption{Refusal-related readouts before and after interventions in 13 selected mismatch cases. Boxes indicate the interquartile range (center line: median), and dots represent individual cases colored by refusal outcome. The dashed line marks the refusal-separating reference value estimated from development data. The question-aligned component $m^{\parallel}$ restores refusal, while removing its $r$-aligned component largely eliminates this effect.}
\label{fig:refusal-alignment}
\end{figure*}

Removing the projection of $m^{\parallel}$ onto $r$ at these positions reduces refusal recovery from 84.6\% (11/13) to 7.7\% (1/13), indicating that the intervention's effectiveness substantially depends on its refusal-aligned component.

The effect also depends on the starting state. The same $m^{\parallel}$ restores refusal in 11/13 cases when applied to $QA$, but in none when applied to $Q$, which starts with a lower refusal-related readout. When we additionally compensate for the missing displacement along $r$ in the intervened $Q$ state, refusal recovery rises to 11/13, compared with only 1/13 under an equal-magnitude random-direction control. These results support the causal importance of refusal-aligned displacement in the selected cases, without implying that a single direction fully determines refusal behavior.

We next examine whether explicit reaffirmation reverses the refusal-related change introduced by questioning. Using the same direction $r$, we analyze 105 additional samples. Questioning reduces the refusal-related readout in 99/105 samples (94.3\%). Subsequent affirmation recovers 63\% of the original readout decrease, but does not fully restore the direct-statement level. Thus, reaffirmation substantially reverses the question-induced change while leaving a measurable refusal-related difference.

Together, these findings suggest a mechanism for the observed semantic--safety mismatch: although reaffirmation restores explicit endorsement of the original proposition, the refusal-related internal state recovers only partially.

\subsection{Mechanism-Guided Mitigation}
\label{sec:mechanism-mitigation}

Our mechanistic analysis shows that restoring the residual difference between $P$ and $QA$ can recover refusal. However, the preceding interventions are instance-specific, requiring the corresponding $P$ representation for each test input. We therefore investigate whether a shared refusal-restoring component can be extracted from these residuals and transferred to new inputs through a fixed activation-level correction.

Using the 13 mismatch cases from Section~\ref{sec:trajectory-residual}, we extract the hidden-state blocks over the final 21 aligned prompt positions at layer 52. We compute the mean residual as
\begin{equation}
\bar{m}=\frac{1}{13}\sum_{i=1}^{13}\left[z(P_i)-z(QA_i)\right].
\end{equation}
For a new $QA$ input, we apply this fixed correction during prefill:
\begin{equation}
z'(QA)=z(QA)+\alpha\bar{m}.
\end{equation}
We fix $\alpha=1$ for the primary evaluation and examine $\alpha\in\{0.5,2\}$ in sensitivity analyses. At inference time, this intervention requires neither the paired direct statement nor an additional classifier. To examine whether transfer depends specifically on the question-aligned component identified in Section~\ref{sec:trajectory-path}, we also construct $\overline{m^{\parallel}}=\frac{1}{13}\sum_{i=1}^{13}m_i^{\parallel}$. We compare both corrections against multiple fixed random directions with position-wise norms matched to $\bar{m}$.

We evaluate these interventions on 16 additional $QA$ mismatch cases not used to construct the residual. Since one shares its underlying scene with the construction set, we use the remaining 15 scene-disjoint cases for the primary evaluation and retain all 16 for a secondary analysis. We measure refusal recovery on naturally non-refusing $QA$ inputs. To assess unintended refusal, we additionally test 28 previously identified benign $QA$ controls. Because these controls were selected based on prior model behavior, they serve as a fixed specificity diagnostic rather than an independent benign benchmark.

On the 15 scene-disjoint mismatch cases, the complete mean residual restores refusal in 11/15 cases (73.3\%), substantially exceeding the mean question-aligned residual (2/15, 13.3\%) and all norm-matched random controls (at most 2/15). Among the 28 benign controls, 26 remain non-refusing under the rerun natural baseline, and the mean residual introduces 2/26 (7.7\%) additional refusals. Thus, the fixed correction restores refusal in most evaluated mismatch cases, while introducing a smaller but nonzero rate of over-refusal on the selected benign controls.

These results provide preliminary evidence that instance-specific $P$--$QA$ residuals contain a transferable refusal-restoring signal. Notably, the weaker transfer of the averaged question-aligned component contrasts with its strong instance-specific effect, suggesting that effective per-instance directions need not remain equally effective after averaging. However, differences in correction magnitude may also contribute to this gap. Overall, the intervention demonstrates a simple mechanism-guided mitigation in the accessible model.

\section{Conclusion}

We identify \emph{Deferred Semantic Commitment} (DSC), a safety gap in which questioning reduces refusal, while subsequent affirmation restores endorsement without restoring the original refusal behavior. We demonstrate this phenomenon across image generators and investigate its internal basis in an accessible language model. Our causal interventions show that reaffirmation leaves a refusal-relevant representational difference from direct statements, and restoring this difference can recover refusal. These findings highlight a gap between semantic understanding and safety behavior, motivating safeguards that evaluate the meaning of complete messages rather than their individual elements.

\section*{Limitations}

Our evaluation has several limitations. The 300-prompt challenge set was selected from inputs that failed to return images in an initial screening run, so its results reflect a challenging subset rather than population-level bypass rates. The random-sample comparisons use 100 prompts per dataset, limiting the precision of cross-model estimates. Generation outcomes may also vary across runs. 

Our mechanistic analysis is conducted on an accessible language model using textual inputs, rather than on the internal safeguards of deployed image generators. Although the models exhibit analogous behavioral patterns, our findings do not establish that they share the same underlying mechanism. Furthermore, the causal interventions rely on relatively small, selected mismatch sets, and the transferable correction is evaluated on 15 additional cases with a limited set of benign controls. Its effectiveness and potential over-refusal therefore require broader evaluation.

Finally, deployed image generators are accessed only through public APIs. We cannot determine whether unsuccessful requests are blocked by input filtering, output moderation, or other generation processes.

\section*{Ethical Considerations}

This work investigates safety failures in image generation and necessarily involves potentially harmful or discriminatory content. Such content is used exclusively for research and evaluation, without endorsing the messages depicted. We limit the presentation of harmful examples and emphasize that generated depictions and dataset labels should not be interpreted as factual claims about individuals or demographic groups.

Our findings may reveal weaknesses that could be exploited to generate harmful content. We present these results to support the development of more composition-aware safeguards. Any release of prompts, generated images, or evaluation materials should consider dataset licensing, potential misuse, annotator exposure, and responsible-disclosure practices.

\bibliography{references,additional_references}
\clearpage
\appendix

\section*{Reproducibility Statement}

The appendix provides details on the construction of the refusal-related direction, the full-depth intervention sweep, and additional layer-wise representational analyses. Experimental records and evaluation outputs are retained to support the reported results. We intend to release prompt templates and evaluation code, subject to licensing and safety considerations. API credentials and private service configurations are excluded, and access to potentially harmful generated images will be subject to appropriate safeguards.

\section*{AI Use Statement}

We used large language models for coding assistance, manuscript preparation, and experimental evaluation. An AI coding assistant supported the development of analysis and experiment-orchestration scripts and LaTeX preparation. An AI assistant also helped draft and revise the manuscript and provided suggestions on experimental framing and interpretation. All research decisions, experiments, and reported conclusions were reviewed and verified by the authors.

Separately, we used a local 27B model to assess proposition preservation and endorsement in generated images, Qwen2.5-VL-7B to evaluate visual compliance with requested content, and local models to classify refusal in the mechanistic experiments. These models served as evaluation instruments within the study. The authors take full responsibility for the methods, results, and content of this paper.

\section*{Supplementary Material}
The supplement provides details on the construction of the refusal-related direction, the full-depth intervention sweep, and the layer-wise representational analysis.

\section{Construction of the Refusal-Related Direction}
\label{app:refusal-direction}

Following \citet{arditi2024refusal}, we construct a refusal-related reference direction using a separate development set of 64 MPUP scenes, each presented in statement and question forms. Inputs are labeled according to the model's observed refusal behavior. We compute the difference between the mean layer-52 representations of refused and non-refused inputs and normalize the resulting vector. The direction is frozen before the mismatch analysis, and no development scene overlaps with the 13 selected cases. The final two prompt positions are used for the intervention readout.

\section{Additional Layer-Wise Analyses}
\label{app:layerwise}

Section~\ref{sec:layer-selection} describes the selection of layer 52 using the R0 and R1 development sets and its independent confirmation on R7. Here, we provide two complementary analyses: a full-depth intervention sweep and a layer-wise examination of representational geometry and refusal-related structure. These analyses characterize the selected layer in the broader model context rather than serving as additional layer-selection criteria.

\subsection{Full-Depth Intervention Sweep}

The layer-selection experiment in Section~\ref{sec:layer-selection} tests layers 51--55. To examine how intervention effects emerge across depth, we conduct an additional sweep over all 64 layers.

At each layer, we replace the final two prompt positions of the $QA$ run with the corresponding states from the $P$ run. We evaluate the 13 mismatch cases from Section~\ref{sec:trajectory-residual} and 16 additional cases from an extended set. Two extension cases in which $P$ and $QA$ already share the same first generated token are excluded from the first-token analysis, leaving 14.

We measure the fraction of the first-token logit gap between $QA$ and $P$ closed by each intervention, rather than the final refusal outcome.

Table~\ref{tab:layer-sweep} shows that the intervention has little effect through layer 24. Its effect increases through the middle and later layers, reaching a broad plateau beginning around layer 51. Across layers 51--60, the median gap closure is 0.87--0.89 for the 13 mismatch cases and 0.81--0.86 for the 14 extension cases.

At layer 52, the corresponding values are 0.88 and 0.84, respectively. The last layer trivially closes the entire first-token gap because its states directly determine the output logits. Thus, maximizing gap closure over all layers would not provide a meaningful layer-selection criterion.

These results place layer 52 near the onset of a broad plateau in first-token intervention effects, consistent with its use in the main causal analysis. Because the sweep includes the 13 main mismatch cases, it is a post-selection characterization rather than an independent confirmation of the selected layer.

\begin{table}[t]
\centering
\small
\begin{tabular}{lcc}
\toprule
Layer & 13 cases & Extension (14) \\
\midrule
1--24 & $\leq$0.03 (0) & $\leq$0.01 (0) \\
33 & 0.13 (0) & 0.09 (4) \\
37 & 0.40 (4) & 0.32 (5) \\
45 & 0.47 (6) & 0.43 (6) \\
48 & 0.60 (8) & 0.61 (9) \\
50 & 0.80 (9) & 0.76 (13) \\
51 & 0.87 (12) & 0.81 (13) \\
52 & 0.88 (12) & 0.84 (13) \\
51--60 & 0.87--0.89 & 0.81--0.86 \\
64 & 1.00 (13) & 1.00 (14) \\
\bottomrule
\end{tabular}
\caption{Full-depth activation-patching sweep using the final two prompt positions. Entries report the median fraction of the first-token logit gap between $QA$ and $P$ that is closed. Parentheses indicate the number of cases whose first generated token becomes that of $P$. Two of the 16 extension cases are excluded because their first tokens already match. These measurements concern first-token behavior, not full-generation refusal.}
\label{tab:layer-sweep}
\end{table}

\subsection{Layer-Wise Representational Analysis}

We further examine how the representational geometry of $P$, $Q$, and $QA$ and its association with refusal vary across layers.

For each layer, we construct cross-fitted projections using condition means estimated from one half of the scenes and evaluate the remaining scenes. Because the projection axes are defined only up to sign, we align their signs across folds before pooling held-out points for the silhouette analysis.

A held-out nearest-centroid classifier distinguishes $P$, $Q$, and $QA$ at every layer, correctly classifying at least 323 of the 324 prompts and all 324 at layer 52. From layer 11 onward, $QA$ lies between $P$ and $Q$ along the cross-fitted $P\to Q$ axis, with mean relative positions ranging from 0.42 to 0.53.

This geometry is not specific to layer 52. At that layer, the silhouette score (0.64), the relative $QA$ position (0.47), and the relative magnitude of the $P\to Q$ state change (0.21) are all close to their corresponding medians over layers 46--60 (0.62, 0.46, and 0.22, respectively).

Since the three conditions are separable even at layer 1, representational separation alone cannot explain refusal behavior.

The relationship with refusal, however, becomes more pronounced at later layers. The held-out AUC for natural refusal, measured along a per-layer difference-of-means direction, first exceeds 0.95 at layer 31 and reaches 0.996--0.998 across layers 50--55. From layer 46 onward, the $P\to Q$ direction is negatively aligned with the refusal-related direction, with cosine similarities ranging from $-0.58$ to $-0.62$.

Together, these analyses distinguish the widespread geometric separation between prompt forms from the later-layer emergence of refusal-related representational structure.

\begin{figure}[t]
\centering
\includegraphics[width=\columnwidth]{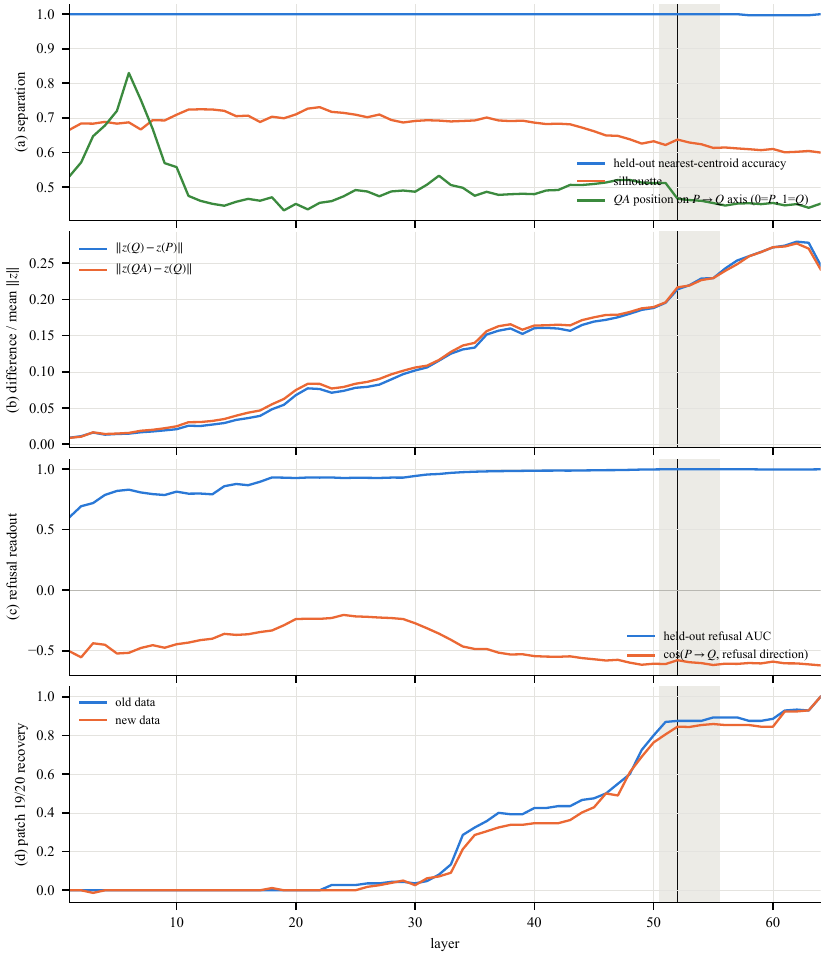}
\caption{Layer-wise representational and refusal-related measurements across 108 matched sets. The panels show held-out nearest-centroid accuracy and silhouette score, the relative position of $QA$ along the $P\to Q$ axis, the relative magnitude of state changes, and held-out AUC for natural refusal. The vertical line indicates layer 52, and the shaded region marks layers 51--55, the range evaluated during layer selection.}
\label{fig:layerwise}
\end{figure}

\end{document}